\documentclass[a4paper,fleqn]{cas-sc}

\usepackage[numbers]{natbib}

\begin{document}
\let\WriteBookmarks\relax
\def\floatpagepagefraction{1}
\def\textpagefraction{.001}

\shorttitle{Structure and dynamics jointly stabilize the international trade hypergraph}

\shortauthors{J.-H Kim et~al.}

\title [mode = title]{Structure and dynamics jointly stabilize the international trade hypergraph}                      



%


\author[1,2]{Jung-Ho Kim}[orcid=0000-0002-1555-4482]
\credit{Data curation, Formal analysis, Investigation, Methodology, Software, Validation, Visualization, Writing – original draft, Writing – review and editing}

\author[3,4]{Sudo Yi}[orcid=0000-0003-2804-7161]
\credit{Data curation, Formal analysis, Investigation, Methodology, Software, Validation, Writing – review and editing}

\author[1,5]{Sang-Hwan Gwak}[orcid=0000-0001-9098-3001]
\credit{Data curation, Formal analysis, Investigation, Methodology, Software, Validation, Writing – review and editing}

\author[1]{K.-I. Goh}[orcid=0000-0003-0385-8208]
\cormark[1]
\ead{kgoh@korea.ac.kr}
\credit{Conceptualization, Data curation, Formal analysis, Funding acquisition, Investigation, Methodology, Project administration, Resources, Supervision, Validation, Writing – review and editing}

\author[3,6]{D.-S. Lee}[orcid=0000-0002-5093-6582]
\cormark[2]
\ead{deoksunlee@kias.re.kr}
\credit{Conceptualization, Data curation, Formal analysis, Funding acquisition, Investigation, Methodology, Project administration, Resources, Supervision, Validation, Writing – original draft, Writing – review and editing}

\affiliation[1]{organization={Department of Physics, Korea University},
    city={Seoul},
    postcode={02841}, 
    country={Korea}
}

\affiliation[2]{organization={Departament d'Enginyeria Inform\`{a}tica i Matem\`{a}tiques, Universitat Rovira i Virgili},
    city={Tarragona},
    postcode={43007}, 
    country={Spain}
}

\affiliation[3]{organization={School of Computational Sciences, Korea Institute for Advanced Study},
    city={Seoul},
    postcode={02455}, 
    country={Korea}
}

\affiliation[4]{organization={CCSS, KI for Grid Modernization, Korea Institute of Energy Technology},
    city={Naju},
    postcode={58330}, 
    state={Jeonnam},
    country={Korea}
}

\affiliation[5]{organization={Large-scale AI research Center, Korea Institute of Science and Technology Information},
    city={Daejeon},
    postcode={34141}, 
    country={Korea}
}

\affiliation[6]{organization={Center for AI and Natural Sciences, Korea Institute for Advanced Study},
    city={Seoul},
    postcode={02455}, 
    country={Korea}
}

\cortext[1]{Corresponding author}
\cortext[2]{Principal corresponding author}


\begin{abstract}
To understand how fluctuations arise and are distributed in international trade, a question crucial for economic risk assessment and policymaking, we analyze strong adverse fluctuations---{\it collapsed trades}---defined as individual trades with sharp annual volume declines.  Adopting a hypergraph framework for a fine-scale trade-centric representation of international trade, we find that collapsed trades (hyperedges) are clustered and their occurrence decays algebraically with trade volume (weight), which suggests inhomogeneous, epidemic-like spreading of collapse in the international trade hypergraph. Modeling collapse propagation as a contagion process and analyzing its dynamics, we show that a positive degree-weight correlation and a volume-decaying collapse rate synergistically suppress the onset of global collective collapse. Notably, the degree-weight correlation persisted but the volume-decay of the collapse rate weakened during the 2008--2009 global economic recession, resulting in a broader collapse spread. Our study shows how the interplay between structure and dynamics stabilizes complex systems.
\end{abstract}




\maketitle

\section{Introduction}
International trade promotes global economic growth but can also transmit economic crises~\cite{Rivera-Batiz2004, 
2002ForbesAre, 
2010KaliFinancial}. During the global economic recession in 2008--2009, the world trade suffered a decline by about $15$ percent, exceeding the decrease in the world gross domestic product (GDP)~\cite{Bems2013}. Even during normal periods, a substantial number of individual trades experience a decline in volume. The formation and distribution of such fluctuations cannot be independent of the organization of international trade~\cite{2021BardosciaThe, 2003SerranoTopology, 2004GarlaschelliFitness-Dependent, 2007HidalgoThe,  2010HeStructure,  2012FronczakStatistical, 2015AlmogA, 2015CiminiSystemic, 2016SaraccoDetecting, 2017BarbierUrn, 2021ChoiSkewness, 2022AlvesThe}. Thus, mapping fluctuations in the international trade network  is fundamentally important,  as it is for general complex systems, and practically relevant for guiding governments and entrepreneurs in designing and optimizing investment and resource allocation across diverse trade partners and industrial sectors~\cite{2010GarasWorldwide, 2011LeeImpact,  2012Brummitt, 2013FotiStability, 2014ContrerasPropagation, 2016LeeStrength, 
2019StarniniThe, 
2020Radicchi, 2021StOnge, 2022KimAnalysis}.

However, practical insights have often been limited, largely because empirical information is not sufficiently incorporated into trade network models. For instance, the organization of international trade has largely been modeled by trade {\it graphs}, which are constructed from either a country-centric or product-centric perspective, offering only a coarse-grained view. Given available data-sets~\cite{2021UNcomtrade} that provide trade volumes for distinct product categories between exporting and importing countries, international trade can be represented at a finer scale by adopting a hypergraph approach, which naturally captures polyadic relationships
~\cite{1989BergeHypergraphs, 2009GhoshalRandom, 2009ZlaticHypergraph, 
2020BattistonNetworks, 2021BattistonThe, 2022MajhiDynamics,2022YiStructure,2025BattistonHOI,2025LernerTemporal}. In the international trade hypergraph (ITH)~\cite{2022YiStructure}, each hyperedge is a  three-vertex set representing an individual trade, connecting two country vertices---one as the exporter and the other as the importer---and a product category vertex, and its weight indicates the corresponding monetary volume of trade. As we show in this study, the hypergraph framework enables detailed characterization of local and global fluctuations, thereby revealing the stabilizing mechanism of international trade.

\begin{figure}[t!]
\centering
\includegraphics[width=\textwidth]{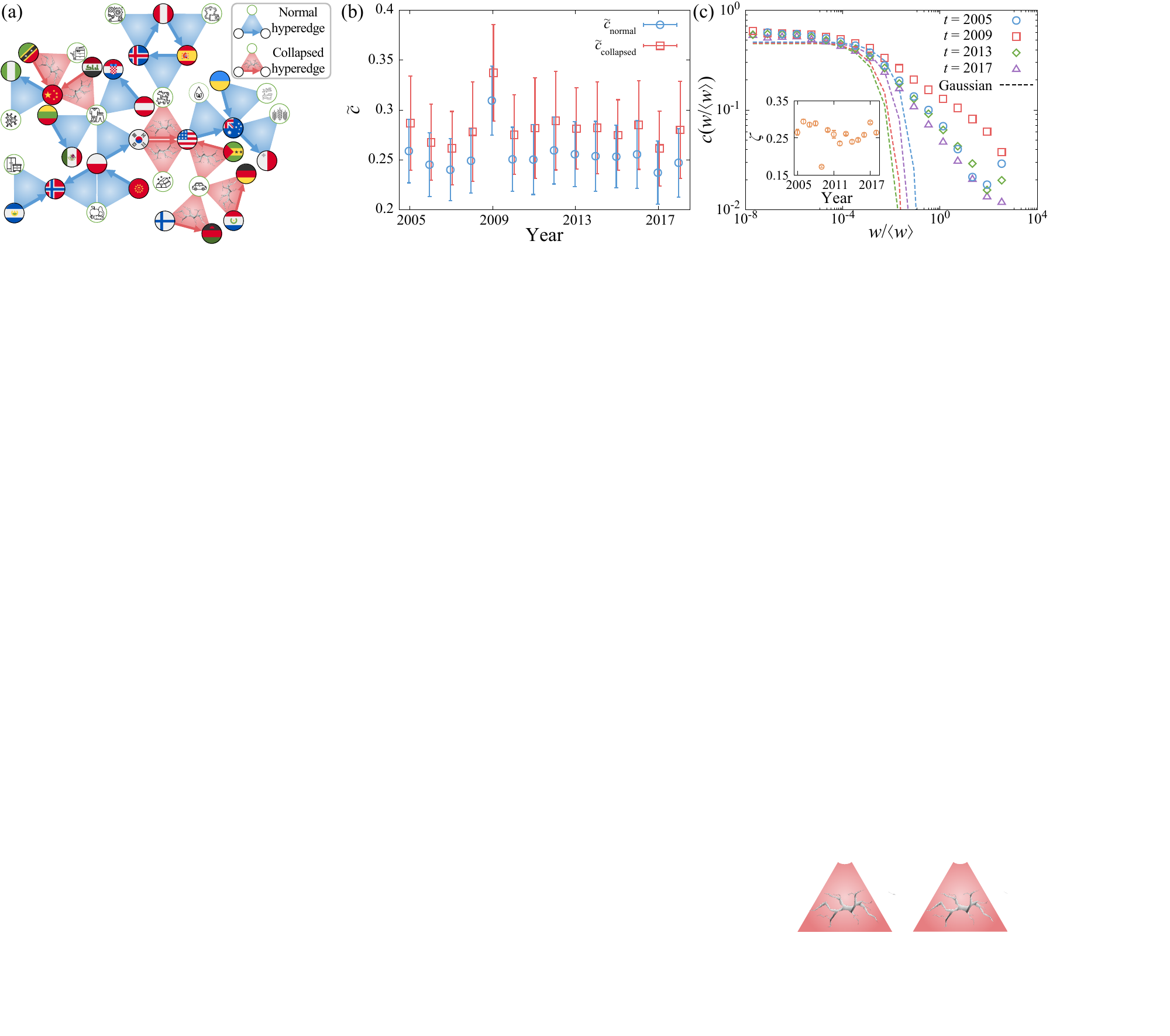}
\caption{Collapsed trades in ITH.
(a) Subhypergraph of ITH in 2009. Hyperedge colors indicate whether the trade is normal or collapsed. Circular vertices contain representative images, with boundary colors denoting vertex types; green for product categories and black for  countries. Arrows represent the direction of product flow.
(b) Proportions of collapsed neighbors around a collapsed ($\tilde{c}_{\text{collapsed}}$) and normal ($\tilde{c}_{\text{normal}}$) hyperedge in empirical data for each year.
Points and error bars indicate mean and standard deviation.
(c) Proportions of collapsed hyperedges, $c(w/\langle w\rangle)$, among hyperedges of given normalized weight for empirical data and for the Gaussian assumption, respectively.
Inset: Power-law decay exponents $\zeta$ in Eq.~\eqref{eq:zeta} estimated from the empirical data by fitting over the range $w / \langle w\rangle=10^{-8} \times 2^{19}$ to $10^{-8} \times 2^{33}$.
Error bars indicate standard error.}
\label{Fig:Empirical}
\end{figure}

Here we define an individual trade as {\it collapsed} if its annual monetary volume in a given year shows a significant decrease compared to the previous year. Mapping collapsed trades onto the ITH reveals that they are clustered, i.e., tend to occur adjacent to one another, suggesting that collapse spreads between neighboring trades.
Moreover, the likelihood of collapse is not uniform across trades but highly inhomogeneous, depending on trade volume. To explore the implications of these features, we introduce and analyze an epidemic-like collapse model to find that collapse propagation in the real-world ITHs is delayed by the combined effect of both structural and dynamical factors: (i) a positive correlation between trade volume and the number of adjacent hyperedges, and (ii) an infection rate decreasing with trade volume. This framework explains the larger collapse during 2008--2009 by a marked weakening of the latter dynamic factor while the former structural factor remained stable. Such interplay of structure and dynamics may serve as a general route to stability in complex systems.


\section{Clustering and volume-dependent incidence of collapsed trades in ITH} 
We construct the ITH [Fig.~\ref{Fig:Empirical}(a)], where vertices represent either countries or product categories (defined by the two-digit code of the standard international trade classification rev. 3~\cite{SITCrev3}) and a hyperedge $h$ represents an individual trade comprising one exporting country, one importing country, and one product category with its weight $w$ given by the annual monetary volume of the corresponding trade. We use trade volume provided by the UN Comtrade database from 2004 to 2018~\cite{2021UNcomtrade} for our analysis.

The organization of these hyperedges is characterized by the adjacency of hyperedges: Two hyperedges are adjacent if they share one or more vertices, as characterized by an integer-weighted adjacency matrix $A_{hh^\prime} = 0,1,2$, or $3$~\footnote{For certain pairs $(h,h')$, we find $A_{hh'}=3$, indicating that they share two countries swapping the roles of an exporter and an importer for the same product.} representing the number of shared vertices~\cite{2024KimHigherOrder}. We define the degree $k_h\equiv \sum_{h^\prime} A_{hh^\prime}$ of hyperedge $h$ as the weighted number of neighboring hyperedges. See Appendix~\ref{seca:datasets} and \ref{seca:basic} for more details of ITH.

\begin{figure}[t!]
\centering
\includegraphics[width=\textwidth]{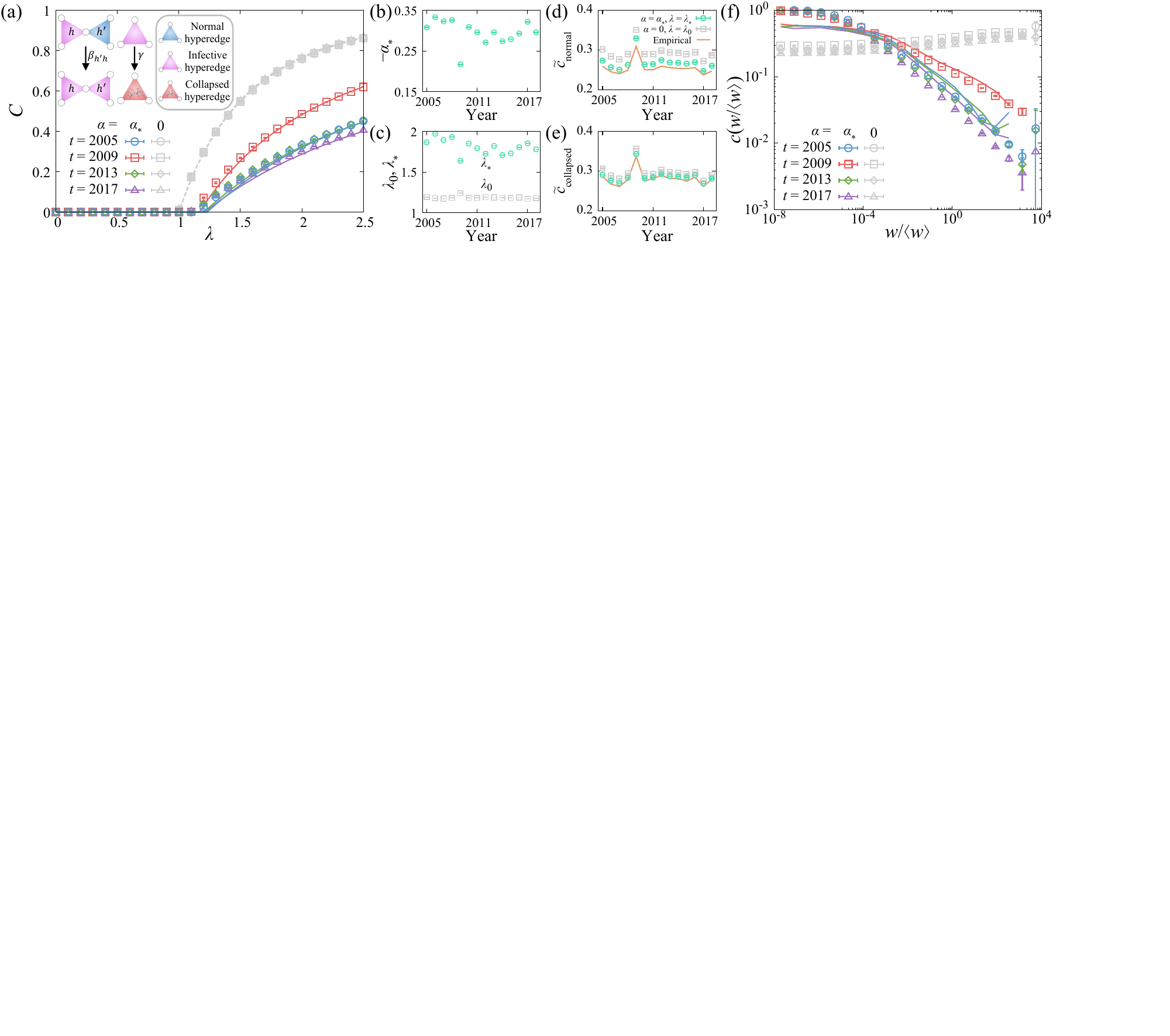}
\caption{Collapse propagation model.
(a) Proportion of collapsed hyperedges in the final state $C$ as a function of $\lambda$ in the collapse propagation model on the empirical ITH with $\alpha=0$ (gray) and $\alpha=\alpha_{*}$ (colored) for selected years $t$. Points represent the Monte Carlo simulation results, and lines represent the mean-field solutions.
(Inset) Schematic illustrations of the transition of a hyperedge $h'$ from $S$ state to $I$ state, due to infection from $h$ with rate $\beta_{h'h}$, and the transition from $I$ state to $C$ state with rate $\gamma$.
(b) $\alpha_{*}$, (c) $\lambda_{0}$ and $\lambda_{*}$ estimated from Monte Carlo simulation results.
(d, e) Mean proportions of collapsed neighbors around (d) a collapsed ($\tilde{c}_{\text{collapsed}}$) and (e) a normal ($\tilde{c}_{\text{normal}}$) hyperedge in Monte Carlo simulation results and empirical data for each year, respectively.
(f) Proportion of collapsed hyperedges, $c(w/\langle w\rangle)$, among the hyperedges of given normalized weight $w/\langle w \rangle$ from Monte Carlo simulation results and empirical data, respectively, for selected years $t$.
Points are the Monte Carlo simulation results, and lines are the empirical data.
Error bars indicate standard error.}
\label{Fig:Model}
\end{figure}


We define a hyperedge $h$ as collapsed if the annual logarithmic change of its weight (trade volume) is sufficiently negative, i.e., 
\begin{equation}
g_h(t) \equiv \log \frac{w_h(t)}{w_h(t-1)}\leq -1,
\label{eq:decline}
\end{equation}
where $w_h(t-1)$ and $w_h(t)$ are the weight in year $t-1$ and $t$, respectively. We find that $g_h(t)$'s are broadly distributed, and about $30\%$ of hyperedges are collapsed, satisfying Eq.~\eqref{eq:decline} each year (see Fig.~\ref{Fig:ITH}). Yet the fraction of collapsed trades, $C$, rises above $0.3$ in 2009 when the global crisis occurred. 

To characterize the topological distribution of these collapsed trades in the ITH,  we measured the average proportions of collapsed neighbors around collapsed ($\tilde{c}_\text{collapsed}$) and normal ($\tilde{c}_\text{normal}$) hyperedges. A significant difference between the two would indicate clustering or dispersion whereas similar values would imply a uniformly random distribution of collapsed trades in the ITH. Figure~\ref{Fig:Empirical}(b) shows that $\tilde{c}_\text{collapsed}>\tilde{c}_\text{normal}$ throughout the observed period indicating that the collapsed hyperedges are more clustered than random. 

Moreover, the likelihood of collapse varies with the weight of a hyperedge. 
The fraction of collapsed hyperedges $c(w/\langle w\rangle)$ 
for given normalized weight $w/\langle w\rangle$ with $\langle \cdot \rangle = N_\text{h}^{-1} \sum_{h} (\cdot)$ denoting the average over all hyperedges decreases algebraically with $w/\langle w\rangle$ as
\begin{equation}
c\left({w}/{\langle w\rangle}\right) \sim \left({w}/{\langle w\rangle}\right)^{-\zeta}
\label{eq:zeta}
\end{equation}
for large $w/\langle w\rangle$ [Fig.~\ref{Fig:Empirical}(c)]. The exponent $\zeta$ stays between 0.24 to 0.29 except for the year 2009 coinciding with the global economic crisis, in which $\zeta \approx 0.17$ [Inset of Fig.~\ref{Fig:Empirical}(c)]. The algebraic decay of collapse probability with volume persists under different thresholds, while $\zeta$ varies across them (see Fig.~\ref{Fig:AlgebraicalDecaying}). Such volume-decay of collapse probability is reminiscent of the  narrower distributions of $g$ for larger-volume trades as also observed in company growth~\cite{stanley_scaling_1996}. 

If we assume that individual trades are composed of micro-contracts whose volumes follow independently Gaussian distributions, then we can expect that the distribution of $g$ is narrower and the collapse probability is reduced for trades of larger volumes (see Appendix~\ref{seca:gaussian} for the derivation). However, the algebraic decay in Eq.~\eqref{eq:zeta} is slower than expected under this Gaussian assumption, showing that the fluctuations of microscopic trades are correlated rather than independent. 

These findings indicate that trade collapses are interconnected and distributed in a highly non-uniform manner. Such  collective and inhomogeneous collapses and global decline may arise from interactions among neighboring trades. If we consider collapse as a disease state, transmission between adjacent hyperedges can reproduce clustered collapses with the volume-dependent collapse probability  incorporated as an inhomogeneous infection rate.


\section{Collapse propagation as a contagion process}
We employ a susceptible-infective-recovered(SIR)-like epidemic model~\cite{2015Pastor-SatorrasEpidemic}, depicted in the inset of Fig.~\ref{Fig:Model}(a), as a hypothetical process of collapse spreading in the ITH. In the model, every hyperedge is in one of three states: normal or susceptible (S), infective (I), and collapsed (C). An infective hyperedge represents a trade currently undergoing collapse (declining in volume) and can infect adjacent susceptible hyperedges---those which have not yet experienced collapse. While trade collapses can arise for various reasons, this model focuses on the transmission of collapse between trades that share countries or products,  driven in real world by factors such as product shortages or deteriorating relations between countries. Each infective hyperedge transitions to the C state at a uniform rate $\gamma$, after which it can no longer transmit collapse. The process continues until no infective hyperedges remain. The outbreak size, defined as the fraction of the collapsed hyperedges in the final state, can then be compared with the empirical fraction of collapsed trades $C$ for each year. This hyperedge-centered model may provide a more refined picture of collapse propagation than the conventional coarse-grained vertex-centered models~\cite{2010GarasWorldwide, 2011LeeImpact}.

The crucial ingredient of our model is a data-inspired inhomogeneous infection rate $\beta_{h'h}$, at which an infective hyperedge $h$ transmits collapse to a susceptible neighbor $h^\prime$ depending on the weight $w_{h'}$ of the target  as [Fig.~\ref{Fig:Model}(a)]
\begin{equation}
\beta_{h^\prime h} = \beta A_{h^\prime h} \frac{w^{\alpha}_{h^\prime}}{\langle w^{\alpha} \rangle}
\label{eq:beta}
\end{equation}
with $\beta$ a characteristic infection rate. The exponent $\alpha$ characterizes a weight-dependent bias in hyperedge infection; with $\alpha<0$, larger-weight hyperedges  have a lower probability of collapse. We set $\beta_{h'h}$ to be independent of the source's weight $w_h$ motivated by the empirical analysis result (Fig.~\ref{Fig:ITH}). Also, we set the infection rate to grow with the number of shared vertices represented by $A_{h'h}$. 

The Monte Carlo simulation of this model on the empirical ITH, starting with a randomly chosen single infective hyperedge each year, shows that the outbreak size $C$ becomes non-zero when the rescaled infection strength $\lambda$, defined by 
\begin{equation}
    \lambda\equiv \frac{\beta}{\gamma}  \frac{\langle k^2\rangle}{\langle k\rangle}
    \label{eq:lambdadef}
\end{equation}
with $\langle k^n\rangle$ the $n$th moment of the degree distribution, 
exceeds a threshold $\lambda_\text{c}$ [Fig.~\ref{Fig:Model}(a)]. Moreover, when $\lambda>\lambda_\text{c}$ and $\alpha$ is negative,  the collapse probability of a hyperedge decreases with its weight algebraically, similar to the empirical observation. 

To obtain simulation results comparable to the empirical data, we estimate $\alpha_*$ and $\lambda_*$  each year such that the outbreak size $C$ and the algebraic-decay exponent  $\zeta$ match closest the corresponding empirical results. The estimated values of $\alpha_*$ are negative [Fig.~\ref{Fig:Model}(b)], consistent with the reduced collapse probability of larger-volume trades. Given that the fraction of collapsed trades are about $0.3$, the values of $\lambda_*$ are larger than the threshold $\lambda_c$ [Fig.~\ref{Fig:Model}(c)]. The Monte Carlo simulations with these estimated  $\alpha_*$ and $\lambda_*$ reproduce excellently  the fractions of collapsed neighbors around a normal and a collapsed hyperedge [Fig.~\ref{Fig:Model}(d,e)].

Using the negative exponent $\alpha_*$ in Eq.~\eqref{eq:beta} is essential for reproducing empirical features.  The outbreak size for given $\lambda$ is smaller with $\alpha_*$ than with $\alpha=0$ [Fig.~\ref{Fig:Model}(a)]. Fixing  $\alpha$ at zero, one can estimate $\lambda_0$ at which the outbreak size matches the empirical fraction of collapsed trades $C$, which is found to be smaller than $\lambda_*$ [Fig.~\ref{Fig:Model}(c)]. In Fig.~\ref{Fig:Model}(d,e), the empirical fraction of collapsed neighbors (lines) is more accurately reproduced by the simulations with $\alpha_{*}$ and $\lambda_{*}$ (colored symbols) than those with $\alpha=0$ and $\lambda=\lambda_{0}$ (gray symbols).  Most remarkably, the collapse probability  depends strongly on $\alpha$; simulations with $\alpha=0$ yield  $c(w/\langle w\rangle)$ that is nearly constant, whereas for $\alpha=\alpha_*$,  it follows the same algebraic decay observed empirically [Fig.~\ref{Fig:Model}(f)]. 

To understand analytically such strong influence of the infection exponent $\alpha$ on collapse spreading, we extend a mean-field approach for inhomogeneous contagion models~\cite{2011BrittonA, 2017MaImportance, 2020NeipelPower-law} to our hypergraph epidemic framework. We will show below that these solutions reveal the role of structural correlations of the ITH in suppressing or promoting the global spread of collapse, in conjunction with the negative $\alpha$.

\section{Synergistic stabilization by structure and dynamics}
Given potential variations of the infection strength $\lambda$ over time, a larger (smaller) outbreak-threshold $\lambda_\text{c}$ indicates that a global spread of collapse is less (more) probable. Therefore, $\lambda_\text{c}$ can quantify the stability of the ITH. Using a mean-field approach~\cite{2011BrittonA, 2017MaImportance, 2020NeipelPower-law} applied to the collapse propagation model on the ITH (Appendix~\ref{seca:mf}),
we find that the probability $\tilde{C}$ that a randomly selected neighboring hyperedge is collapsed in the final state satisfies $\tilde{C} = \left\langle {k \over \langle k\rangle} \left(1 - e^{-{\beta\over \gamma} k {w^\alpha\over \langle w^\alpha\rangle}\tilde{C}}\right)\right\rangle$ (see Appendixes), from which we obtain the threshold $\lambda_\text{c}$ 
\begin{equation}
\lambda_\text{c} \equiv \frac{ \langle k^2\rangle \langle w^\alpha\rangle}{\langle k^2w^\alpha\rangle},
\label{eq:lambdac}
\end{equation}
and the outbreak size $C$ 
shown in Fig.~\ref{Fig:Model}(a). As $\alpha$ takes on larger negative values, it not only suppresses the outbreak size but also increases the threshold $\lambda_\text{c}$ for the empirical ITH. Therefore, we may consider the smaller value of $|\alpha_{*}|$ in 2009 compared to other years [Fig.~\ref{Fig:Model}(b)] as a hallmark of the low stability.

\begin{figure}[t!]
\centering
\includegraphics[width=0.5\textwidth]{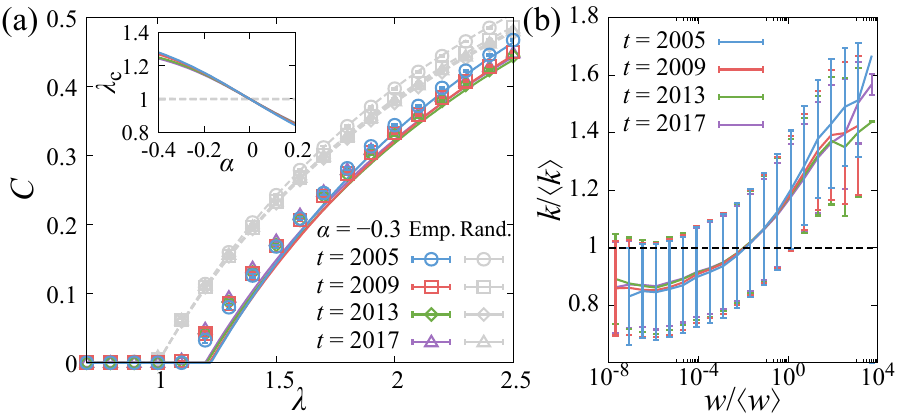}
\caption{Effects of dynamical inhomogeneity and structural correlation.
(a) Outbreak size $C$ versus rescaled infection strength $\lambda$ on the empirical (Emp.) and randomized (Rand.) ITHs with $\alpha$ fixed at $-0.3$.
Inset: Critical point $\lambda_\text{c}$ as a function of $\alpha$ on the empirical and randomized ITH for each year.
Points are the Monte Carlo simulation results, and  lines are the mean-field results.
Error bars indicate standard error.
(b) Mean normalized degree $k/\langle k \rangle$ of hyperedges of given normalized weight $w/\langle w \rangle$ in empirical data for each year.
The dashed line represents the case where there is no correlation between hyperedge's degree and weight.
Error bars indicate standard deviation.}
\label{Fig:Correlation}
\end{figure}

These findings raise a natural question: Why does a negative $\alpha$ enhance the stability of the ITH?  Equation~\eqref{eq:lambdac} provides the answer. Rearranging Eq.~\eqref{eq:lambdac}, we obtain 
\begin{equation}
\frac{1}{\lambda_\text{c}}-1 = \frac{\langle (k^2 - \langle k^2\rangle)(w^\alpha - \langle w^\alpha\rangle ) \rangle}{\langle k^2 \rangle \langle w^\alpha \rangle},
\label{eq:lambdac2}
\end{equation}
which includes a normalized covariance of $k^2$ and $w^\alpha$, revealing the effect of structural correlation. This shows that $\lambda_\text{c}=1$ if $\alpha=0$ or there is no degree-weight correlation.  When $\alpha<0$, as implicated empirically, a positive (negative) correlation between $k$ and $w$ makes the normalized covariance of $k^2$ and $w^\alpha$ in Eq.~\eqref{eq:lambdac2} negative (positive), thereby increasing (decreasing) $\lambda_\text{c}$. To examine such crucial influence of structural correlations, we construct randomized ITHs 
by randomly shuffling the weights of hyperedges in the empirical ITHs. For these randomized ITHs, $\lambda_\text{c}$ is fixed at $1$, independent of $\alpha$ [inset of Fig.~\ref{Fig:Correlation}(a)]. In contrast, in the empirical ITHs, the degree of hyperedges increases with their volume  [Fig.~\ref{Fig:Correlation}(b)], indicating a positive degree-weight correlation similar to that observed in trade networks~\cite{2008BhattacharyaThe, 2009FagioloWorld-trade}. This correlation raises
$\lambda_{\text{c}}$, and reduces outbreak sizes for given $\lambda$, compared with the randomized ITHs [Fig.~\ref{Fig:Correlation}(a)]. While the volume dependence of the collapse probability weakened, the positive degree-weight correlation persisted even in 2009 during the global crisis [Fig.~\ref{Fig:Correlation}(b)].

These results suggest that a negative dynamical correlation between collapse probability and weight ($\alpha<0$) and a positive structural correlation between degree and weight synergistically serve to mitigate the spread of collapse from local to global scales. This mechanism can be understood intuitively as follows.
A local collapse is most likely to occur in hyperedges of small weight due to $\alpha<0$. Because these hyperedges are likely to have small degrees under the positive degree-weight correlation, the local collapse spreads to only a few neighbors, reducing the reproduction number $R_0=\lambda/\lambda_\text{c}$ compared to  ITHs without a degree-weight correlation or a negative $\alpha$. A change in either the negative dynamical or positive structural correlation can destabilize the ITH. We have shown that the global crisis in 2009 can be attributed to the weakening of the dynamical correlation, while the structural correlations remained robust. Furthermore, if degree and weight were negatively correlated, the volume-decaying collapse probability would promote global collapse spreading. Therefore, our study reveals a joint stabilization mechanism, driven by the interplay between structure and dynamics in the world trade system.

\section{Discussion}
Here we have studied fluctuations in international trade, focusing on the significant decline in trade volume, which we refer to as `collapse', at the level of individual trades. Taking a hypergraph approach, we found that such collapsed trades (hyperedges) are not uniformly distributed but tend to be clustered in the ITH. Furthermore, the proportion of collapsed trades decays algebraically with their volume (weight). To investigate the implications of these findings for the emergence of global-scale collapse, we introduced an effective model of  collapse spreading with an inhomogeneous infection rate, motivated by the key empirical features that we identified. Using the mean-field theory and Monte Carlo simulations, we have shown that the combined effects of the structural correlation between hyperedge degree and weight and the dynamical correlation between infection rate and volume can either facilitate or suppress global collapse spreading,  depending on their signs. Most importantly, the real-world ITH exhibits a positive degree-weight correlation and a collapse probability decreasing with hyperedge weight, which together synergistically suppress the global outbreak of collapse. Given various forms of correlations identified in diverse real-world complex networks,  our findings highlight  how such correlations, both structural and dynamical, can jointly serve to regulate fluctuation spreading and promote stability. 

The origins of the dynamical  and structural correlations in the ITH remain to be fully understood. Future research could explore the correlations between microscopic transactions, which may explain why the proportion of collapsed trade decays more slowly than expected if these transactions were independent. Datasets capturing microscopic trades on shorter time scales could also provide a more detailed information on the collapse dynamics in the ITH~\cite{2022DiemQuantifying, 2024ChakrabortyInequality}. Additionally, analyzing extended datasets over longer time periods may reveal how the structure of the ITH evolves, shedding light on the evolutionary origins of the structural correlation.

\printcredits

\section*{Declaration of competing interest}
The authors declare that they have no known competing financial interests or personal relationships that could have appeared to influence the work reported in this paper.

\section*{Acknowledgments}
This work was supported in part by the National Research Foundation of Korea (NRF) grant funded by the Korea government (MSIT) [No.~RS-2025-00558837 (K.-I.G.), No.~NRF-2019R1A2C1003486 (D.-S.L.), No.~RS-2023-00279802 and No.~RS-2022-NR071795 (S.Y.)], by KIAS Individual Grants [No.~CG074102 (S.Y.) and No.~CG079902 (D.-S.L.)] at Korea Institute for Advanced Study, and by the KENTECH Research Grant [No. KRG-2021-01-007 (S.Y.)]. 

\section*{Data availability}
The data that support the findings of this study are available from UN Comtrade, but their public availability is restricted as they were used under license for the current study. The source codes can be freely accessed at the GitHub repository: \url{https://github.com/JungHoKim4823/InternationalTradeHypergraph2025}.

\appendix
\numberwithin{equation}{section} 
\numberwithin{figure}{section}
\section{Datasets}
\label{seca:datasets}
The international trade instances in the UN Comtrade database are derived from the reports of importing countries on imported products, with volume measured in US dollars, along with information about the partner exporters. Countries reporting at least one instance of import and the product categories defined by the two-digit code of the standard international trade classification rev. 3 are included in the ITH.

\section{Basic features of ITH}
\label{seca:basic}
\begin{figure}[t]
\centering
\includegraphics[width=\textwidth]{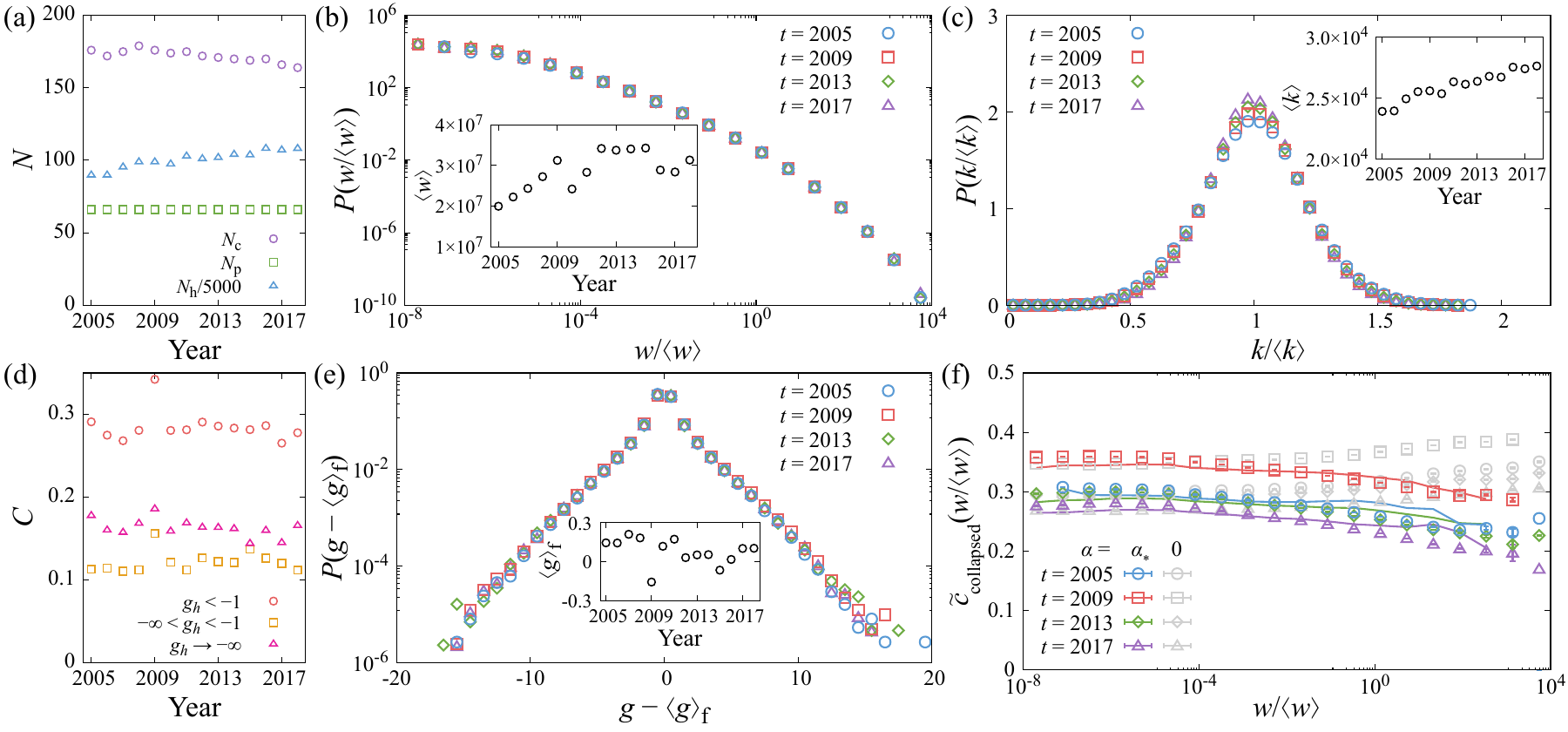}
\caption{Basic features of ITH.
(a) Numbers of countries ($N_{\text{c}}$), product categories ($N_{\text{p}}$), and hyperedges ($N_{\text{h}}$) in the ITH for each year.
(b)  Distribution of the normalized weight ${w}/{\langle w\rangle}$ with the mean  $\langle w\rangle = N_\text{h}^{-1} \sum_h w_h$ for each year.
Inset: Mean weight $\langle w \rangle$ versus time. 
(c) Distribution of the normalized degree ${k}/{\langle k\rangle}$ with the mean $\langle k\rangle = N_\text{h}^{-1} \sum_h k_h$ for each year.
Inset: Mean degree $\langle k \rangle$ versus time. 
(d) Proportion of collapsed trades under various thresholds for defining collapse in Eq.~\eqref{eq:decline}.
(e) Distribution of the logarithmic change of trade volume $g$ shifted by the mean $\langle g\rangle_{\text{f}}$ for each year. The subscript `f' denotes averaging over hyperedges exhibiting finite $g$.
Inset: Mean logarithmic change $\langle g\rangle_{\text{f}}$ versus time. 
(f) The mean proportion $\tilde{c}_{\text{collapsed}}({w}/{\langle w\rangle})$ of collapsed neighbors around a collapsed hyperedge of given normalized weight ${w}/{\langle w\rangle}$.
Error bars indicate standard error.}
\label{Fig:ITH}
\end{figure}

The ITH for year $t$ is built using the annual trade volumes $\{w_h(t-1)\}$ from the previous year and thus represents a snapshot as of January 1 of year $t$. During year $t$, the ITH evolves through the appearance or disappearance of hyperedges or vertices, as well as fluctuations in hyperedge weights. These changes are reflected in the annual trade volumes $\{w_h(t)\}$ recorded as of December 31 of year $t$, which can be compared with $\{w_h(t-1)\}$. The set $\{w_h(t)\}$ is then used to construct the ITH for year $t+1$.

Throughout the observed period, the numbers of countries and products (product categories) remain nearly constant at about $170$ and $66$ respectively [Fig.~\ref{Fig:ITH}(a)]. The  number of hyperedges increases slightly from $450,000$ to $540,000$ implying a world-wide growing activity in international trade. The volume of individual trade is quite heterogeneous as revealed by the broad distributions displaying fat tails that show a good collapse when trade volume is normalized by the mean, i.e., ${w}/{\langle w \rangle}$~[Fig.~\ref{Fig:ITH}(b)]. The mean weight $\langle w\rangle = N_\text{h}^{-1} \sum_h w_h$ tends to increase over time [inset of Fig.~\ref{Fig:ITH}(b)].
On the other hand, the normalized degree ${k}/{\langle k \rangle}$ of individual trade follows a narrow bell-shaped distribution [Fig.~\ref{Fig:ITH}(c)].
The mean degree $\langle k\rangle = N_\text{h}^{-1} \sum_h k_h$ increases over time, meaning that the ITH becomes denser [inset of Fig.~\ref{Fig:ITH}(c)].

We find the fraction of collapsed trades $C \approx 0.3$ under the criterion in Eq.~\eqref{eq:decline}~as shown in Fig.~\ref{Fig:ITH}(d), with its ephemeral increase in 2009 when the global crisis occurred. The distribution of the annual logarithmic change of hyperedge weight $g$ is broad, exhibiting exponential-like tails and ranging between $-20$ to $20$ across years [Fig.~\ref{Fig:ITH}(e)]. In 2009, the mean value of $g$ decreased due to the global economic recession.

The mean proportion $\tilde{c}_{\text{collapsed}}({w}/{\langle w\rangle})$ of collapsed neighbors around a collapsed hyperedge of  normalized weight ${w}/{\langle w\rangle}$ is nearly independent of ${w}/{\langle w \rangle}$ [Fig.~\ref{Fig:ITH}(f)], allowing us to neglect the source-weight dependence when formulating the infection rate in Eq.~\eqref{eq:beta} of the collapse propagation model.

\section{Robustness of algebraic volume-decay of collapse probability}
\label{seca:decay}
\begin{figure}[t]
\centering
\includegraphics[width=\textwidth]{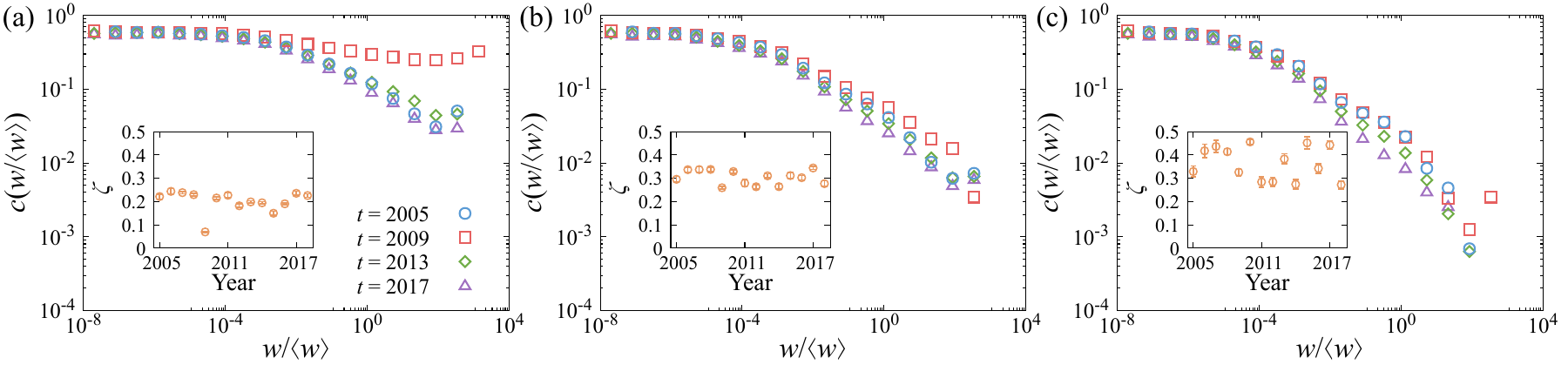}
\caption{Proportion of collapsed hyperedges, $c(w / \langle w\rangle)$, among the hyperedges of given normalized weight $w / \langle w\rangle$ in empirical data.  The collapse threshold in Eq.~\eqref{eq:decline}, set to $-1$ in the main text, is here varied as (a) $-0.5$, (b) $-2$, and (c) $-\infty$.
Insets: The estimated power-law decay exponents $\zeta$ in the empirical data.}
\label{Fig:AlgebraicalDecaying}
\end{figure}

As shown in Fig.~\ref{Fig:AlgebraicalDecaying}, the algebraic decay of collapse probability with trade volume remains robust, although the exponent $\zeta$ varies, when the collapse threshold in Eq.~\eqref{eq:decline} is varied as $-0.5$, $-2$, and $-\infty$.

\section{Gaussian assumption}
\label{seca:gaussian}
In this section we assume that each trade in year $t$ consists of multiple microscopic transactions of volume $w_{0}=\$ 1,000$ each and that from year $t$ to $t+1$, the volumes of those microscopic transaction change by $\Delta w_0$'s that follow a Gaussian distribution independently.  Under this {\it Gaussian} assumption, we derive the distribution of the annual change $\Delta w$ of a (macroscopic) trade, and compute the probability that $\Delta w$ satisfies Eq.~\eqref{eq:decline}, yielding the fraction of collapsed trades with given normalized weight $c^{\rm (Gaussian)}(w/\langle w\rangle)$, which can be compared with the empirical data in Eq.~\eqref{eq:zeta}. 

The mean $\langle \Delta w_{0} \rangle$ and standard deviation $\sigma_{\Delta w_{0}}$ of the annual volume change $\Delta w_{0}$  for independent microscopic transactions are approximated by the mean and standard deviation of the annual volume changes,from year $t$ to $t+1$, of trades with volumes between \$950 and \$1,050 in each year $t$ in the empirical data. We then assume that $\Delta w_0$ follows a Gaussian distribution $P_0^{\rm (Gaussian)}(\Delta w_0) = {1\over \sqrt{2\pi \sigma_{\Delta w_{0}}^2}} \exp\left[{-{(\Delta w_0 -\langle \Delta w_{0} \rangle )^2 \over 2\sigma_{\Delta w_{0}}^2 }}\right]$. Under this assumption, a macroscopic trade of volume $w$ consists of $N=\lceil w/w_{0} \rceil$ independent microscopic transactions, so its annual volume change is the sum of  $N$ independent $\Delta w_0$'s. 
Consequently, the annual change $\Delta w$ from $w$ to ($w+\Delta w$) of the considered trade follows a Gaussian distribution with mean $N \langle \Delta w_{0} \rangle$ and standard deviation $N^{1/2} \sigma_{\Delta w_{0}}$, $P^{\rm (Gaussian)}(\Delta w) = {1\over \sqrt{2N\pi \sigma_{\Delta w_{0}}^2}} \exp\left[{-{(\Delta w -N\langle \Delta w_{0} \rangle )^2 \over 2N\sigma_{\Delta w_{0}}^2 }}\right]$. According to Eq.~\eqref{eq:decline}, a trade of volume $w$ in year $t$ is classified as collapsed if its annual change $\Delta w$ is smaller than $-(1-e^{-1})w$  as collapsed.  Therefore  the collapse probability $c^{\rm (Gaussian)}(w/\langle w\rangle)$ of normalized weight $w/\langle w\rangle$ is given by 
\begin{equation}
c^{\rm (Gaussian)}(x)
= \int_{-\infty}^{-(1-e^{-1})x\langle w\rangle} P^{\rm (Gaussian)}(\Delta w)=\frac{1}{\sqrt{2\pi N \sigma^2_{\Delta w_0}}} \int_{-\infty}^{-(1-e^{-1})x\langle w \rangle}d(\Delta w) \exp\left[{-\frac{(\Delta w - N \langle \Delta w_{0} \rangle)^2}{2N \sigma_{\Delta w_0}^2}}\right].
\end{equation}

\section{Monte Carlo simulation}
\label{seca:mc}
Using a parallel update and a discrete time scheme with $\gamma=1$, the Monte Carlo simulation of the collapse propagation model proceeds as follows. i) Initially all hyperedges are susceptible except for a randomly-selected one, which is infective. 
ii) At every discrete time step, we traverse all the infective hyperedges $h$ to determine whether to infect each of their susceptible neighbor hyperedges $h'$ with probability $\beta_{h'h}$ given in Eq.~\eqref{eq:beta}. After this, all the previously infective hyperedges collapse---except for the newly infected ones---as we set $\gamma=1$. The process ii) continues until no infective hyperedges are left. We then measure the fraction $C$ of the collapsed hyperedges in the final state, ranging between $0$ and $1$. To avoid finite-size effects, we perform 100 simulations and use only those realizations that exhibit $C >10^{-3} \approx N_\text{h}^{-1/2}$ with $N_\text{h}$ the total number of hyperedges.

\section{Mean-field theory for the collapse propagation model}
\label{seca:mf}
The most remarkable characteristics of the collapse propagation model is the inhomogeneous infection rate given in Eq.~\eqref{eq:beta}. To understand how this dynamical inhomogeneity affects collapse propagation in the ITH,  we consider the probabilities $s_h, i_h, c_h$ of a hyperedge $h$ to be susceptible, infective, and collapsed, respectively, which evolve with time as 
\begin{align}
{d s_{h} \over dt} &= -k_h \langle i_{h'} \beta_{hh'}\rangle_{\text{nn}(h)} s_h = -k_h \beta {w_h^\alpha \over \langle  w^\alpha \rangle} \tilde{i} \, s_h,
\label{AP_S0_v2}\\
{d i_{h} \over dt} &= k_h \beta {w_h^\alpha \over \langle  w^\alpha \rangle} \tilde{i}\, s_h-\gamma i_{h},\label{AP_I0_v2}\\
{d c_{h} \over dt} &= \gamma \,i_{h},\label{AP_R0_v2}
\end{align}
in which $\langle f\rangle \equiv {\sum_h f_h \over \sum_h 1}$ is the global average of the property $f$ of a hyperedge and $\langle f\rangle_{\text{nn}(h)} \equiv  {\sum_{h'} A_{h'h} f_{h'}\over \sum_{h'} A_{h'h}}$ is the local-neighbor average of $f$, the average over the neighboring hyperedges of a given hyperedge $h$.  Here we adopt the mean-field approximation, replacing the local-neighbor average $\langle f\rangle_{\text{nn}(h)}$ by the global-neighbor average $\tilde{f} \equiv {\sum_{hh'} A_{hh'} f_{h'} \over \sum_{hh'} A_{hh'}}={1\over N_\text{h}}\sum_{h' } {k_{h'}\over  \langle k\rangle} f_{h'}$ with $\langle k\rangle = N^{-1}_\text{h}\sum_h k_h$ representing the mean degree of hyperedges.  Notice that the global-neighbor average $\tilde{i}$ can be considered as the probability that a neighbor hyperedge of a randomly-selected hyperedge is infective. The probability  $c_h$ does not decrease with time and we are interested in its value in the final state $C = \lim_{t\to\infty}{1\over N_\text{h}}\sum_h c_h(t)$, which can be compared to the empirical fraction of collapsed hyperedges $C$. 

Introducing the probability of a neighbor hyperedge (of a randomly-selected hyperedge) to be collapsed $\tilde{c} \equiv {1\over N_\text{h}}\sum_h {k_h \over \langle k\rangle}c_h$, we find immediately from Eq.~\eqref{AP_R0_v2} that 
\begin{equation}
{d\tilde{c}\over dt} = \gamma \tilde{i}.
\label{eq:dtildec}
\end{equation}  
From Eqs.~\eqref{AP_S0_v2} and ~\eqref{eq:dtildec}, we obtain the probability $s_h$ as a function of $\tilde{c}$ as
\begin{equation}
s_h(\tilde{c})=   e^{-{\beta \over \gamma} k_h {w^\alpha_h \over \langle w^\alpha\rangle}\tilde{c}}
\label{eq:sh}
\end{equation}
with $s_h(0)=1$ used.   The time-evolution of $\tilde{i}$ is obtained by averaging Eq.~\eqref{AP_I0_v2} over all  hyperedges weighted by their degrees and using Eq.~\eqref{eq:sh} as 
\begin{equation}
{d\tilde{i}\over d\tilde{c}} =  -1+ {1\over N_\text{h}}\sum_h \left\{{k_h^2 \over \langle k\rangle} {\beta \over \gamma} {w^\alpha_h \over \langle w^\alpha\rangle} s_h(\tilde{c})\right\}.
\label{eq:dtildei}
\end{equation}
Analyzing Eqs.~\eqref{eq:dtildec}, \eqref{eq:sh}, and \eqref{eq:dtildei}, one can obtain the condition for the global outbreak of collapse and its magnitude. Below, we derive the solutions in two limiting cases: the early-time regime and  the final state. 

\subsection{Early-time regime}
In the early-time regime when ${\beta \over \gamma} k_h {w^\alpha_h \over \langle w^\alpha\rangle}\tilde{c}\ll 1$ for all $h$, all hyperedges are likely to be normal such that $s_h(\tilde{c})\simeq s_h(0)= 1$ and therefore Eq.~\eqref{eq:dtildei} reads as  ${d\tilde{i}\over d\tilde{c}}\simeq-1+ {\beta \over \gamma} {\langle k^2 w^\alpha \rangle \over \langle k\rangle\langle w^\alpha\rangle}   =-1 + R_0$, in which the reproduction number $R_0$ is given by 
\begin{equation}
R_0 \equiv {\beta \over \gamma} {\langle k^2 w^\alpha\rangle \over \langle k\rangle \langle w^\alpha\rangle}= \lambda {\langle k^2 w^\alpha \rangle \over \langle k^2\rangle\langle w^\alpha\rangle}
\label{eq:R_0}
\end{equation}
with the rescaled infection strength $\lambda$ in Eq.~\eqref{eq:lambdadef}~used. Notice that the rescaled infection strength $\lambda$ is equivalent to the reproduction number $R_0$ in uncorrelated ITHs - where the degree and weight of a hyperedge are independent such that $\langle k^2w^\alpha\rangle = \langle k^2 \rangle \langle w^\alpha\rangle$. This motivated us to use $\lambda$ as a dimensionless characteristic infection strength throughout this study. In the early-time regime, we thus obtain $\tilde{i} \simeq (R_0-1)\tilde{c}$, which is inserted into Eq.~\eqref{eq:dtildec} to give ${d\tilde{i}\over dt} = \gamma \tilde{i} (R_0-1)$, leading to 
\begin{equation}
\tilde{i}\sim e^{\gamma(R_0-1)t}.
\label{eq:tildei}
\end{equation}
This reveals that the initial cluster of infective hyperedges can grow if the reproduction number $R_0$ is larger than one, as is well known in the compartmental models for epidemic spreading, or equivalently if in our model the rescaled infection strength $\lambda$ exceeds a threshold $\lambda_\text{c}$
\begin{equation}
 \lambda>\lambda_\text{c} \equiv {\langle k^2\rangle\langle w^\alpha\rangle\over \langle k^2 w^\alpha \rangle }.
\label{eq:supercritical}
\end{equation}
We note that the influence of a hyperedge's degree in Eqs.~\eqref{eq:dtildei} and \eqref{eq:R_0} appears quadratically rather than linearly. This arises because a hyperedge $h$ of degree $k_h$ can both be infected at a rate proportional to $k_h$  [Eq.~\eqref{AP_I0_v2}] and transmit infection to a number of susceptible neighbors also proportional to $k_h$. 

\subsection{Final state} 
Integrating Eq.~\eqref{eq:dtildei}, we obtain
\begin{equation}
\tilde{i}(\tilde{c}) = -\tilde{c} +{1\over N_\text{h}}\sum_h {k_h \over \langle k\rangle}  \left[1 - e^{-{\beta\over \gamma} k_h {w^\alpha_h\over \langle w^\alpha\rangle}\tilde{c}}\right]   
\end{equation}
with $\tilde{i}(0)=0$ used approximately. In the final state, no infective hyperedges remain, i.e., $\tilde{i}=0$ and the corresponding value of $\tilde{c}$ in equilibrium, which we denote by $\tilde{C}$, satisfies 
\begin{equation}
\tilde{C} = {1\over N_\text{h}}\sum_h {k_h \over \langle k\rangle} \left[1 - e^{-{\beta\over \gamma} k_h {w^\alpha_h\over \langle w^\alpha\rangle}\tilde{C}}\right].
\label{eq:tildecinf}
\end{equation}
The right-hand-side of Eq.~\eqref{eq:tildecinf} is a concave function of $\tilde{C}$ that increases monotonically  with increasing $\tilde{C}$, starting from $0$ at $\tilde{C}=0$ and converging to $1$ as $\tilde{C} \to \infty$. In addition to the trivial solution $\tilde{C}=0$ to Eq.~\eqref{eq:tildecinf}, we find a positive (and stable) solution  when Eq.~\eqref{eq:supercritical} holds, i.e., $\lambda>\lambda_\text{c}$.  For $\lambda<\lambda_\text{c}$, the trivial solution ($\tilde{C}=0$) is the only and stable solution.  Moreover, when $\lambda$ is slightly larger than $\lambda_\text{c}$, one can obtain approximately $\tilde{C}$ by expanding the right-hand-side of Eq.~\eqref{eq:tildecinf} for small $\tilde{C}$ as 
\begin{equation}
\tilde{C} \simeq 
2 \left({\lambda \over \lambda_\text{c}}-1\right) {\langle k^2w^\alpha\rangle^2 \over \langle k\rangle \langle k^3 w^{2\alpha}\rangle}.
\end{equation}

Plugging the solution $\tilde{C}$ of Eq.~\eqref{eq:tildecinf} into Eq.~\eqref{eq:sh}, we obtain  $s_h(\tilde{C})$, which denotes the probability of a  hyperedge to remain susceptible in the final state. Therefore the fraction of collapsed hyperedges  in the final (equilibrium) state is evaluated as 
\begin{equation}
C = {1\over N_\text{h}}\sum_h \left\{1 - s_h(\tilde{C})\right\} 
= {1\over N_\text{h}}\sum_h \left(1 -e^{-{\beta \over \gamma} k_h {w^\alpha_h \over \langle w^\alpha\rangle}\tilde{C}} \right).
\label{eq:cinf}
\end{equation}
From the behavior of $\tilde{C}$ as a function of $\lambda$, we find that $C$ is zero for $\lambda<\lambda_\text{c}$ and positive for $\lambda>\lambda_\text{c}$. Moreover, when $\lambda$ is larger than but near $\lambda_\text{c}$, we find 
\begin{equation}
C \simeq 
 2 (\lambda - \lambda_\text{c}) {\langle k w^\alpha\rangle \langle k^2 w^\alpha \rangle^2 \over \langle k^2\rangle \langle w^\alpha\rangle \langle k^3 w^{2\alpha}\rangle}.
\end{equation}

\subsection{Relation between $\alpha$ and $\zeta$} 
Finally, let us consider the probability $c(w/\langle w\rangle)$ of an individual hyperedge of given normalized weight $w/\langle w\rangle$ to be collapsed. Rewriting Eq.~\eqref{eq:cinf}  as $C = \sum_{w/\langle w\rangle} P(w/\langle w\rangle) c(w/\langle w\rangle)$ with $P(w/\langle w\rangle)=N_\text{h}^{-1}\sum_h \delta(w_h/\langle w\rangle - w/\langle w\rangle)$ denoting the distribution of normalized weight, we obtain

\begin{align}
c\left(\frac{w}{\langle w\rangle}=x\right) &= {\sum_h \left(1 -  e^{-{\beta \over \gamma} k_h {w^\alpha_h \over \langle w^\alpha\rangle}\tilde{C}}\right)\delta(w_h -x\langle w\rangle)\over \sum_h \delta(w_h-x\langle w\rangle)} \nonumber \\
&\simeq 2  (\lambda - \lambda_\text{c}) x^\alpha \frac{\langle k(x)\rangle}{\langle k\rangle} \frac{ \langle k\rangle\langle w\rangle^\alpha \langle k^2 w^\alpha\rangle^2}{\langle k^2\rangle \langle w^\alpha\rangle \langle k^3 w^{2\alpha}\rangle}, 
\end{align}
where the last approximation holds for $\lambda$ slightly larger than $\lambda_\text{c}$ and $\langle k(w/\langle w\rangle=x)\rangle$ is the conditional mean degree of the hyperedges of normalized weight $w/\langle w\rangle$ given by $\langle k(x)\rangle= {\sum_h k_h \delta(w_h-x\langle w\rangle)\over \sum_h \delta(w_h -x\langle w\rangle)}$. Therefore, we find when $\lambda$ is slightly larger than $\lambda_\text{c}$ that the volume-dependency of collapse probability is given by 
\begin{equation}
c\left(\frac{w}{\langle w\rangle}=x\right)\propto x^\alpha \, \frac{\langle k(x)\rangle}{\langle k\rangle}.
\label{eq:cx}
\end{equation} 
We have estimated $\alpha_*$ such that the empirical observation $c(w/\langle w\rangle)\sim (w/\langle w\rangle)^{-\zeta}$ in Eq.~\eqref{eq:zeta} is best reproduced in the Monte Carlo simulations.  Equation~\eqref{eq:cx} indicates that if degree and weight were not correlated in the ITH, i.e., $\langle k\rangle(w/\langle w\rangle)$ were independent of $w/\langle w\rangle$, then $c(w/\langle w\rangle)\sim (w/\langle w\rangle)^{\alpha_*}$, i.e., $\zeta = -\alpha_*$. 
However, as seen by comparing inset of Fig.~\ref{Fig:Empirical}(c) and Fig.~\ref{Fig:Model}(b), $\zeta$ is slightly smaller than $-\alpha_*$ throughout the studied period. This difference can be attributed to the positive degree-weight correlation in the real-world ITH shown in Fig.~\ref{Fig:Correlation}(b), which is expected to follow $\langle k(w/\langle w\rangle)\rangle \sim (w/\langle w\rangle)^{|\alpha_*|-\zeta}$ according to Eq.~\eqref{eq:cx}.

\bibliographystyle{model1-num-names}

\bibliography{TradeHypergraph}

\end{document}